\documentclass[conference]{IEEEtran}

\ifCLASSINFOpdf
\else
\fi
\usepackage{multicol}
\usepackage{tikz}
\usepackage{tkz-graph}
\usepackage{bbm}
\usepackage{booktabs}
\usepackage{comment}
\usepackage{listings}

\usepackage{caption}
\usepackage{algpseudocode}
\usepackage{algorithm}
\usepackage[english]{babel}

\usepackage{color}
\usepackage[cmex10]{amsmath}
\usepackage{dblfloatfix}
\usepackage{pgfplots}
\usepackage{pgfplotstable}
\usepackage{hyperref}

\addto\extrasenglish{%

}

\begin{document}
\pagenumbering{gobble}

%
\title{Process Mining for Python (\textbf{PM4Py}): Bridging the Gap Between Process- and Data Science}

\author{
\IEEEauthorblockN{Alessandro Berti\IEEEauthorrefmark{1}, Sebastiaan J. van Zelst\IEEEauthorrefmark{1}\IEEEauthorrefmark{2}, Wil M.P. van der Aalst\IEEEauthorrefmark{1}\IEEEauthorrefmark{2}}
\IEEEauthorblockA{\IEEEauthorrefmark{1}RWTH Aachen University\\
Process and Data Science group, Lehrstuhl f\"ur Informatik 9, 52074 Aachen, Germany\\
\{a.berti,s.j.v.zelst,wvdaalst\}@pads.rwth-aachen.de}
\IEEEauthorblockA{\IEEEauthorrefmark{2}Fraunhofer Gesellschaft\\
Institute for Applied Information Technology (FIT), Sankt Augustin, Germany\\
\{sebastiaan.van.zelst,wil.van.der.aalst\}@fit.fraunhofer.de
}}

\maketitle

\begin{abstract}
Process mining, i.e., a sub-field of data science focusing on the analysis of event data generated during the execution of (business) processes, has seen a tremendous change over the past two decades.
Starting off in the early 2000's, with limited to no tool support, nowadays, several software tools, i.e., both open-source, e.g., ProM and Apromore, and commercial, e.g., Disco, Celonis, ProcessGold, etc., exist.
The commercial process mining tools provide limited support for implementing custom algorithms. Moreover, both commercial and open-source process mining tools are often only accessible through a graphical user interface, which hampers their usage in large-scale experimental settings.
Initiatives such as RapidProM provide process mining support in the scientific workflow-based data science suite RapidMiner. However, these offer limited to no support for algorithmic customization.
In the light of the aforementioned, in this paper, we present a novel process mining library, i.e. Process Mining for Python (PM4Py) that aims to bridge this gap, providing integration with state-of-the-art data science libraries, e.g., pandas, numpy, scipy and scikit-learn.
We provide a global overview of the architecture and functionality of PM4Py, accompanied by some representative examples of its usage.
\end{abstract}




%
\IEEEpeerreviewmaketitle

\section{Introduction}
The field of \emph{process mining}~\cite{DBLP:books/sp/Aalst16} provides tools and techniques to increase the overall knowledge of a (business) process, by means of analyzing the event data stored during the execution of the process.
Process mining received a lot of attention from both academia and industry, which led to the development of several commercial and open-source process mining tools.
The majority of these tools supports \emph{process discovery}, i.e., discovering a process model that accurately describes the process under study, as captured within the analyzed event data.
However, process mining also comprises \emph{conformance checking}, i.e., checking to what degree a given process model is accurately describing event data, and \emph{process enhancement}, i.e., techniques that enhance process models by projecting interesting information, e.g. case flow and/or performance measures, on top of a model.
The support of such types of process mining analysis is typically limited to open source, academic process mining tools such as the ProM Framework~\cite{van2005prom} and Apromore~\cite{la2011apromore}.

Both ProM and Apromore put a significant emphasis on non-expert usability, i.e., by means of providing an easy to use graphical user interface.
Whereas such an interface helps to engage non-expert users and, furthermore, helps to showcase process mining to a larger audience, it hampers the usability of the tools for the purpose of large-scale scientific experimentation \cite{bolt2016scientific}.
To this end, the RapidProM~\cite{mans2014supporting,DBLP:journals/corr/AalstBZ17} initiative allows for repeated execution of large-scale experiments with process mining algorithms in the RapidMiner\footnote{http://rapidminer.com} suite.
However, RapidProM provides neither easy algorithmic customization nor an easy way to integrate custom developed algorithms.
As such, the aforementioned tools fail to support customizable process mining algorithms and large-scale experimentation and analysis.

\begin{figure*}[tb]
\centering
\begin{lstlisting}[language=Python, basicstyle=\footnotesize, frame=single, numbers=left]
from pm4py.algo.discovery.alpha import versions
from pm4py.objects.conversion.log import factory as log_conversion
ALPHA_VERSION_CLASSIC = 'classic'
ALPHA_VERSION_PLUS = 'plus'
VERSIONS = {ALPHA_VERSION_CLASSIC: versions.classic.apply,
ALPHA_VERSION_PLUS: versions.plus.apply}
def apply(log, parameters=None, variant=ALPHA_VERSION_CLASSIC):
    return VERSIONS[variant](log_conversion.apply(log, parameters, log_conversion.TO_EVENT_LOG), parameters)
\end{lstlisting}
\caption{Example factory method (Alpha Miner). Different variants (the Alpha and the Alpha+) are made available.}
\label{fig:factoryMethod}
\end{figure*}

To bridge the aforementioned gap, i.e., the lack of process mining software that i) is easily extendable, ii) allows for algorithmic customization and iii) allows us to easily conduct large scale experiments, we propose the \emph{Process Mining for Python (PM4Py)} framework.
To achieve the aforementioned goals, a fresh look on the currently available programming languages and libraries indicates that the Python programming language\footnote{ http://python.org}, along with its ecosystem, is most suitable.
In particular, the data science world, both for classic data science (pandas, numpy, scipy \ldots) and for cutting-edge machine learning research (tensorflow, keras \ldots), is heavily using Python.
Other libraries, albeit with a lower number of features, exist already for the Python language (PMLAB \cite{carmona2014pmlab}, OpyenXES \cite{valdiviesoopyenxes}). The bupaR library \cite{janssenswillen2017bupar} supports process mining in the statistical language R, that is widely used in data science.
The main focal points of the novel PM4Py library are:
\begin{itemize}
\item Lowering the barrier for algorithmic development and customization when performing a process mining analysis compared to existing academic tools such as ProM \cite{van2005prom}, RapidProM \cite{mans2014supporting} and Apromore \cite{la2011apromore}.
\item Allow for the easy integration of process mining algorithms with algorithms from other data science fields, implemented in various state-of-the-art Python packages.
\item Create a collaborative eco-system that easily allows researchers and practitioners to share valuable code and results with the process mining world.
\item Provide accurate user-support by means of a rich body of documentation on the process mining techniques made available in the library.
\item Algorithmic stability by means of rigorous testing.
\end{itemize}

The remainder of this paper is structured as follows.
In \autoref{sec:architecture}, we present the architecture and an overview of the features provided by PM4Py. 
In \autoref{sec:features}, we present some representative examples (process discovery, conformance checking).
\autoref{sec:maturity} discusses the maturity of the tool and \autoref{sec:conclusion} concludes this paper.

\section{Architecture and Features}
\label{sec:architecture}

In order to maximize the possibility to understand and re-use the code, and to be able to execute large-scale experiments, the following architectural guidelines have been adopted on the development of PM4Py:
\begin{itemize}
\item A strict separation between \emph{objects} (event logs, Petri nets, DFGs, \ldots), \emph{algorithms} (Alpha Miner \cite{van2004workflow}, Inductive Miner \cite{leemans2015scalable}, alignments \cite{adriansyah2011cost} \ldots) and \emph{visualizations} in different packages.
In the {\it pm4py.object} package, classes to import/export and to store the information related to the objects are provided, along with some utilities to convert objects (e.g. process trees into Petri nets); while in the {\it pm4py.algo} package, algorithms to discover, perform conformance checking, enhancement and evaluation are provided.
All visualizations of objects are provided in the {\it pm4py.visualization} package.
\item Most functionality in PM4Py has been realized through \emph{factory methods}.
These factory methods provide a single access point for each algorithm, with a standardized set of input objects, e.g., event data and a parameters object.
Consider the factory method of the Alpha Miner, depicted in \autoref{fig:factoryMethod}. 
The Alpha \texttt{(variant='classic')} and the Alpha+ \texttt{(variant='plus')} are made available.
Factory methods allow for the extension of existing algorithms whilst ensuring backward-compatibility.
The factory methods typically accept the name of the variant of the algorithm to use, and some parameters (shared among variants, or variant-specific). 
\end{itemize}

In the remainder of this section, we present the main features of the library, organized in objects, algorithms, and visualizations.

\subsection{Object Management}
Within process mining, the main source of data are \emph{event data}, often referred to as an \emph{event log}.
Such an event log, represents a collection of events, describing what activities have been performed for different instances of the process under study.
PM4Py provides support for different types of event data structures:
\begin{itemize}
\item \emph{Event logs}, i.e., representing a list of \emph{traces}. 
Each trace, in turn, is a list of events.
The events are structured as key-value maps.
\item \emph{Event Streams} representing one list of events (again represented as key-value maps) that are not (yet) organized in cases.
\end{itemize}
Conversion utilities are provided to convert event data objects from one format to the other.
Furthermore, PM4Py supports the use of \texttt{pandas data frames}, which are efficient in case of using larger event data.
Other objects currently supported by PM4Py include: heuristic nets, accepting Petri nets, process trees and transition systems.

\begin{figure*}[tb]
\centering
\begin{lstlisting}[language=Python, basicstyle=\footnotesize, frame=single, numbers=left]
from pm4py.objects.log.importer.xes import factory as xes_importer
from pm4py.algo.discovery.alpha import factory as alpha_miner
from pm4py.visualization.petrinet import factory as pn_vis_factory
log = xes_importer.apply("C:\\receipt.xes")
# discovers a Petri net along with an initial (im)
# and a final marking (fm)
net, im, fm = alpha_miner.apply(log)
gviz = pn_vis_factory.apply(net, im, fm)
pn_vis_factory.view(gviz)
\end{lstlisting}
\caption{PM4Py code to load a log, apply Alpha Miner and visualize a Petri net.}
\label{fig:examplePm4py}
\end{figure*}

\begin{figure*}[!b]
\centering
\begin{lstlisting}[language=Python, basicstyle=\footnotesize, frame=single, numbers=left]
from pm4py.algo.conformance.alignments import factory as alignments
# alignments accepts a log and an accepting Petri net, i.e.
# a Petri net along with an initial (im) and a final (fm) marking
aligned_traces = alignments.apply(log, net, im, fm)
for index, result in enumerate(aligned_traces):
    print(index, result['alignment'])
\end{lstlisting}
\begin{lstlisting}[language=Python, basicstyle=\footnotesize, frame=single]
[('register request', 'register request'), ('>>', None), ('check ticket', 'check ticket'),
('examine thoroughly', 'examine thoroughly'), ('>>', None), ('decide', 'decide'), ('>>', None),
('reject request', 'reject request')]
\end{lstlisting}
\caption{PM4Py code to perform alignments between a log and a model, and print the alignments. The output of the alignment of a trace on an example log and model is reported.}
\label{fig:alignments}
\end{figure*}

\subsection{Algorithms}
The PM4Py library provides several mainstream process mining techniques, including:
\begin{itemize}
\item \emph{Process discovery}: Alpha(+) Miner \cite{van2004workflow} and Inductive Miner (IMDF \cite{leemans2015scalable}).
\item \emph{Conformance Checking}: Token-based replay and alignments \cite{adriansyah2011cost}.
\item Measurement of fitness, precision, generalization and simplicity of process models.
\item Filtering based on time-frame, case performance, trace endpoints, trace variants, attributes, and paths.
\item Case management: statistics on variants and cases.
\item Graphs: case duration, events per time, distribution of a numeric attribute's values.
\item Social Network Analysis \cite{van2004mining}: handover of work, working together, subcontracting and similar activities networks.
\end{itemize}

\subsection{Visualizations}
\begin{figure}[tb]
\centering
\includegraphics[width=240px]{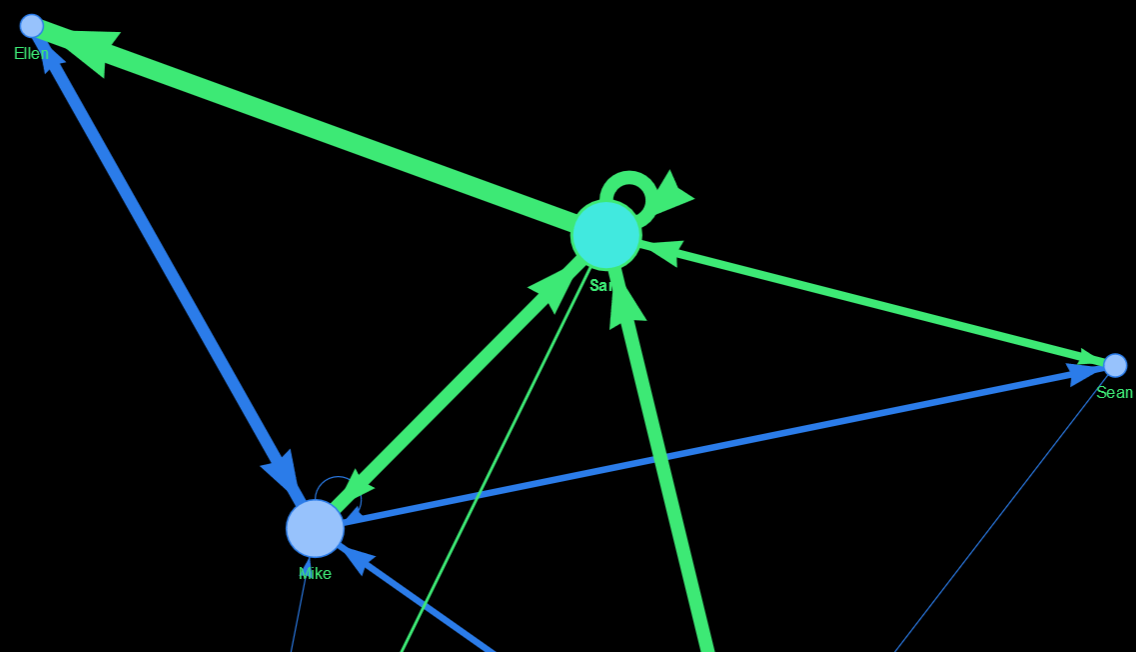}
\caption{Social Network Analysis (Handover of Work metric) using Pyvis visualization.}
\label{fig:sna}
\end{figure}

The following Python visualization libraries have been used in the project:
\begin{itemize}
\item GraphViz: representation of directly-follows graphs, Petri nets, transition systems, process trees.
\item NetworkX: static representation of social networks.
\item Pyvis: web-based, dynamic representation of social networks (see \autoref{fig:sna}).
\end{itemize}

\section{Examples}
\label{sec:features}
In this section, we provide some examples of the use of PM4Py.

\subsection{Process Discovery}
\begin{figure}[tb]
\centering
\includegraphics[width=200px]{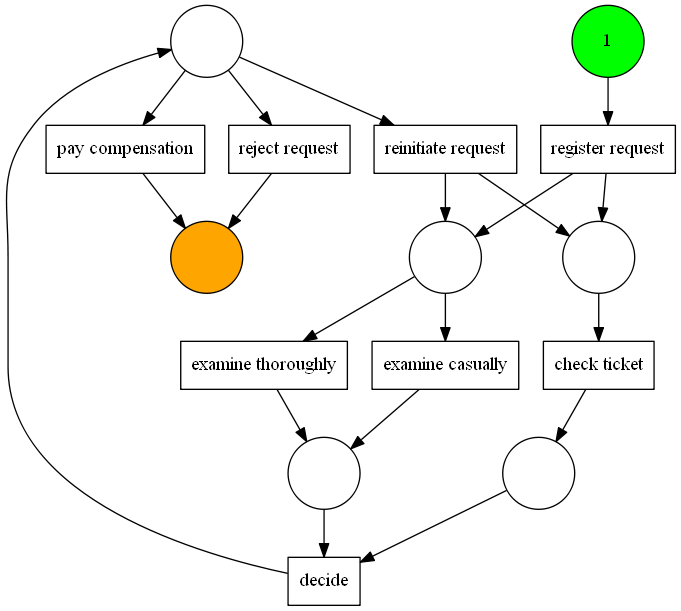}
\caption{PM4Py in action: process discovery with the Alpha Miner.}
\label{fig:alphaminer}
\end{figure}

\autoref{fig:examplePm4py} shows example code to perform process discovery using Alpha Miner and visualize the process model.
The factory methods that are needed (XES importer, Alpha Miner and Petri net visualization) are loaded (line 1-3). 
Then, an XES log is imported (line 4), the Alpha Miner is applied providing the log object (line 7), and the visualization is obtained: a factory method is applied to layout the graph (line 8), and the result is shown in a window (line 9). 
The result is shown in \autoref{fig:alphaminer}.

\subsection{Conformance Checking}

\autoref{fig:alignments} shows example code to apply alignments and display the result. 
First, the alignments factory method is loaded (line 1). 
Then, the alignments between a log object and a process model are obtained (line 4). 
For each aligned trace (line 5) the alignment result is displayed on the screen (line 6). 
The alignment of a trace is reported in the lower part of \autoref{fig:alignments}.

\section{Maturity of the tool}
\label{sec:maturity}

\begin{figure}[tb]
\centering
\includegraphics[width=\columnwidth]{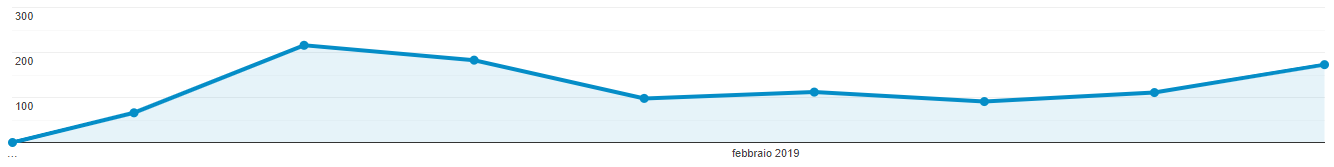}
\caption{Users that accessed the PM4Py website in February 2019}
\label{fig:usersWebsite}
\end{figure}

\begin{figure}[tb]
\centering
\includegraphics[width=\columnwidth]{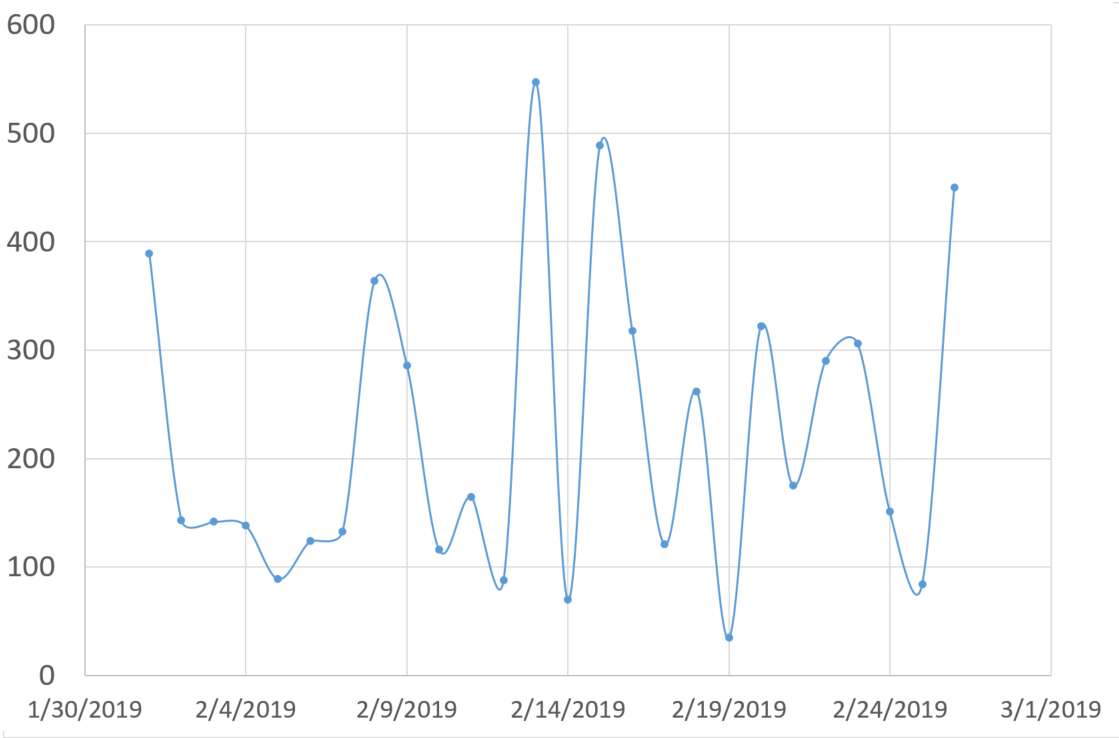}
\caption{Daily downloads of PM4Py from Pypi during the month of February 2019}
\label{fig:pepytech}
\end{figure}

PM4Py 1.0 has been released on 21/12/2018 and was used by $200$ students in the ``Introduction to Data Science'' course held by the
Process and Data Science group in the RWTH Aachen University.
Already two academic projects have been supported by PM4Py and are publicly available:
\begin{itemize}
\item Usage of probabilistic automata for compliance checking ({\it https://github.com/lvzheqi/StreamingEventCompliance}).
\item Prefix alignments for streaming event data \cite{van2017online} ({\it https://gitlab.com/prefal/confo}).
\end{itemize}
PM4Py 1.1 has been released on 22/02/2019 with additional features. There are some integrations
of the PM4Py library in other projects:
\begin{itemize}
\item bupaR R process mining library uses PM4Py to handle alignments and get models using the Inductive Miner.
\item A data analytics web interface was written in Vue.JS ({\it https://git.bogdan.co/b0gdan/beratungsleistungen}).
\end{itemize}
In \autoref{fig:usersWebsite}, some statistics taken from Google Analytics are reported about the number of accesses to PM4Py web site during the month of February 2019.
In \autoref{fig:pepytech}, some statistics about the downloads of the PM4Py library from PIP are reported.
Issues are managed through Github.
The XES certification, with maximum score, has been awarded to the PM4Py library.

\section{Conclusion}
\label{sec:conclusion}



In this paper, the PM4Py process mining library has been introduced.
PM4Py supports a rapidly growing set of process mining techniques (discovery, conformance checking, enhancement \ldots).
A video presenting the library and some example applications (log management, process discovery, conformance checking) has been made available\footnote{http://pm4py.pads.rwth-aachen.de/pm4py-demo-video/}.
The library can be installed\footnote{Additional prerequisites, available at the page http://pm4py.pads.rwth-aachen.de/installation/ have to be installed.} through the command {\it pip install pm4py}.
Extensive documentation is provided through the official website of the library. 
Moreover, the Github repository supports a collaborative eco-system where users could signal problems or contribute to the code.

\bibliographystyle{IEEEtran}
\bibliography{pm4pygeneral}

\end{document}